\documentclass[%
 reprint,
 amsmath,amssymb,
 aps,
]{revtex4-1}

\usepackage{graphicx}
\usepackage{dcolumn}
\usepackage{bm}
\usepackage{siunitx}
\usepackage{textcomp}
\usepackage{xcolor}

\begin{document}

\title{Quantum decoherence by Coulomb interaction}

\author{N. Kerker$^1$, R. R\"{o}pke$^1$, L.M. Steinert$^1$, A. Pooch$^1$  and A. Stibor$^{1,2*}$}
\affiliation{\mbox{$^1$Institute of Physics and Center for Collective Quantum Phenomena in LISA$^+$, University of T\"{u}bingen, T\"{u}bingen, Germany} \mbox{$^2$Lawrence Berkeley National Lab, Molecular Foundry, Berkeley, USA, *email: astibor@lbl.gov}}

\begin{abstract}
The performance of modern quantum devices in communication, metrology or microscopy relies on the quantum-classical interaction which is generally described by the theory of decoherence. Despite the high relevance for long coherence times in quantum electronics, decoherence mechanisms mediated by the Coulomb force are not well understood yet and several competing theoretical models exist. Here, we present an experimental study of the Coulomb-induced decoherence of free electrons in a superposition state in a biprism electron interferometer close to a semiconducting and metallic surface. The decoherence was determined through a contrast loss at different beam path separations, surface distances and conductibilities. To clarify the current literature discussion, four theoretical models were compared to our data. We could rule out three of them and got good agreement with a theory based on macroscopic quantum electrodynamics. The results will enable the determination and minimization of specific decoherence channels in the design of novel quantum instruments. 
\end{abstract}

\maketitle 

{\it Introduction.--} A deep understanding of the transition region between a quantum and a classical system is necessary for various fields of physics and technical applications. Decoherence, being the loss of quantum behaviour is described by a variety of theoretical approaches \cite{Zeh1970,Zurek1981,Joos2003,Zurek2003,Anglin1997,Scheel2012,Machnikowski2006,Howie2011}. The question of how complex and massive an object can possibly be to still show quantum mechanical behavior like superposition \cite{Gerlich2011,Arndt2014} and how strong the influence of gravitational \cite{Pikovski2015} or electromagnetic \cite{Sonnentag2007} interactions with a "measuring" environment is, induced numerous experiments in the last decades \cite{Hornberger2003,Hackermuller2004,Gerlich2007}. At the same time, quantum hybrid systems appeared in several realizations \cite{Wallquist2009}, like photons with cavity mirrors \cite{Chan2011}, ultracold atoms near nanotubes \cite{Gierling2011}, cantilevers \cite{Hunger2010} or microwave cavities \cite{Rabl2006}, indicating the rise of new devices that work in this transition regime. New ideas in quantum electron microscopy \cite{Putnam2009,Kruit2016} and quantum information science \cite{Roepke2019} were proposed and identified to be restricted by the decoherence of the electrons in a superposition state. These systems have in common that a quantum object interacts with a classical microscopic solid state device, with promising applications in quantum information science \cite{Nielsen2010}, quantum metrology \cite{Giovannetti2006} and quantum simulation \cite{Georgescu2014}. To enhance such technologies, it is desired not only to understand the quantum-classical transition, but also to control the interaction of a quantum system with its environment and to keep the quantum state coherent as long as possible. 
A good approach to study this field is to take a rather simple, well understood quantum system and gradually turn on decoherence by introducing more and more interaction with the classical environment. This can be performed by e.g.~the emission of photons out of Rydberg atoms in a microwave cavity \cite{Brune1996} or out of helium atoms after interacting with a standing light wave \cite{Pfau1994}. Also, decoherence studies with neutron matter waves crossing oscillating magnetic fields have been conducted \cite{Sulyok2012,Sulyok2010}. Matter wave interferometers are in particular suitable for such experiments, where decoherent interactions lead to a gradually leakage of which-path information and a deteriorating fringe contrast. This has been performed for C$_{70}$ fullerene molecules, were laser excitation \cite{Hackermuller2004} or collisions with gas atoms \cite{Hornberger2003} led to decoherence.

In all these cases the quantum object was neutral. To determine the role of decoherence by Coulomb interaction, an experiment was proposed by Anglin et al.~\cite{Anglin1996,Anglin1997} and performed by Sonnentag et al.~\cite{Sonnentag2005,Sonnentag2007} where electrons are interfered in an biprism interferometer. The separated and coherent matter wave paths pass aloof a semiconducting silicon surface and interact with the electron gas inside by the Coulomb force. The decoherence was measured in dependence of the path separation and the electron beam distance to the surface. The theoretical analysis of the results could satisfactorily describe the distribution, but the strength of the decoherence had to be adjusted by fit factors that deviated significantly from the predicted values \cite{Sonnentag2007,Anglin1997,Machnikowski2006}. 

\begin{figure*}
\includegraphics[width=7in]{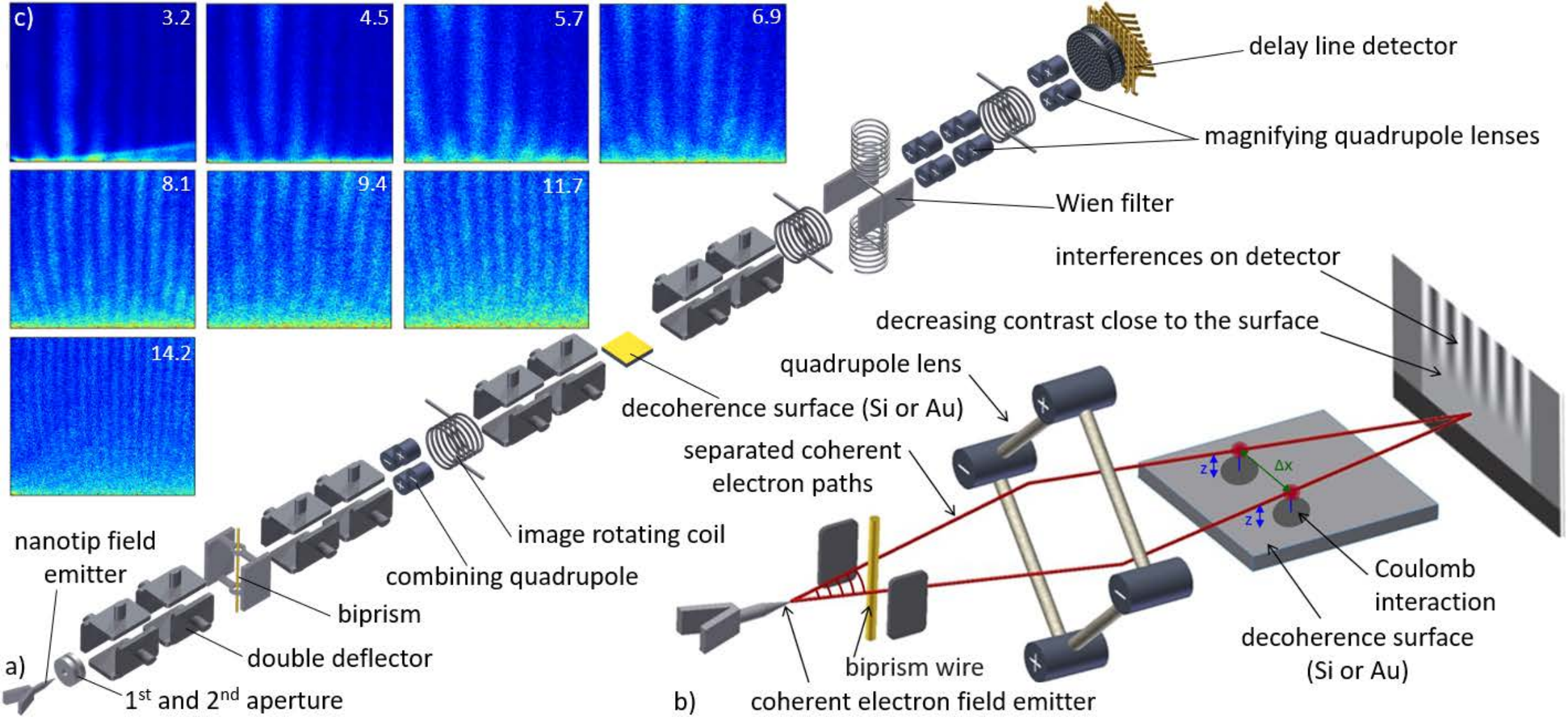} \caption{(Color online) a) Experimental setup for electron decoherence measurements close to a semiconducting and metallic environment. b) Sketch of the experiment, where a coherent electron beam is separated and combined by an electrostatic biprism and a quadrupole lens. In the quantum superposition state before interference, the beam passes aloof a semiconducting n-doped silicon or gold surface, interacting with the bulk electron gas. This process causes decoherence which can be observed in a decreasing interference contrast close to the surface. c) Recorded fringe pattern above the silicon surface (bottom edge of each image) for different beam path separations $\Delta x$ (value in upper right corner in \textmu m). The top edge is at a height $z=$ \SI{40}{\micro\metre}. An increase in signal above the surface up to max.~\SI{4}{\micro\metre} is visible that is assumed to be due to scattering on waver edge defects. } \label{fig1}
\end{figure*}

In this article, we present a Coulomb-induced decoherence measurement that is in good agreement to a current decoherence theory by Scheel et al.~\cite{Scheel2012} without the need of a fit factor. Similar to the experiment of Sonnentag et al.~\cite{Sonnentag2007}, the electrons are prepared in a superposition state close to a semiconducting doped silicon surface in a biprism interferometer. We additionally performed measurements aloof of a gold surface. The theoretical approach of Scheel et al.~relies on macroscopic quantum electrodynamics based on linear-response theory \cite{Scheel2012}. Our data agree well with this theory in a Markov approximation. Additionally, our experiment with silicon is able to rule out the finite temperature approximation of Scheel et al.~\cite{Scheel2012} and three other current decoherence approaches by Anglin et al.~\cite{Anglin1997}, Machnikowski~\cite{Machnikowski2006} and Howie \cite{Howie2011}. For gold, the decoherence is significantly weaker due to the high surface conductibility. Our data is also in favor of the Scheel theory but resolution needs to be improved for a final conclusion. This study provides a better understanding of the complex mechanisms that change a quantum to a classical system. It is relevant for future technological applications in quantum electron microscopy, soft-surface analysis and quantum information science.

{\it Setup and procedure.--} The decoherence measurements are performed in an electron biprism interferometer with a large beam path separation \cite{Mollenstedt1956,Hasselbach1988}. Fig.~\ref{fig1}~a) provides the experimental configuration of all implemented electron optical components and the sketch in Fig.~\ref{fig1}~b) illustrates the principle of the measurement. A significant part of the setup was resumed from the experiment by Sonnentag et al.~\cite{Sonnentag2007}. A coherent and monochromatic electron beam is generated with an energy of \SI{1}{keV} by a single atom nano-tip field emitter \cite{Kuo2006,Kuo2008} close to two extraction apertures \cite{Pooch2018}.
The electron waves are guided and aligned on the optical axis by four double deflectors. The beam is separated by an electrostatic biprism wire \cite{Schuetz2014} on a negative potential and combined again with a quadrupole lens. After superposition, the resulting matter wave interference pattern is amplified by further quadrupole lenses and visualized by a delay line detector \cite{Jagutzki2002}.
The potential gradient between the deflector electrodes can cause an interference contrast loss due to the limited longitudinal coherence length. This well-known effect is corrected by a Wien filter consisting of two deflector plates and two magnetic coils \cite{Nicklaus1993}.
The electron beam path separation was varied by adapting the biprism and first quadrupole lens voltages which also changes the superposition angle and the fringe distance. The total beam size was limited by \SI{1}{mm} apertures and the diameter of the coherent overlapping before magnification was around \SI{5}{\micro\meter} at the beam path separation of \SI{9.4}{\micro\meter} in fig.~\ref{fig1} c). To measure the Coulomb-induced decoherence of the superposition state, the separated coherent electron paths are send aloof and parallel to a \SI{1}{\centi\m} long semiconducting plate of doped silicon that is placed behind the first quadrupole lens. In a separate experiment a gold plate was introduced as the decoherence surface. Its resistivity (\SI{2.2e-6}{\ohm\cm} \cite{Kittel1996}) is six orders of magnitude smaller than the one of the n-doped silicon (\SI{1.5}{\ohm\cm}), causing a significant difference in the decoherence behaviour. The silicon or gold plate can be moved up and down normal to the surface by a micrometer stage to calibrate the horizontal $z$ distance between the electrons and the plate by its shadow on the detector after magnification. Three imaging rotating coils allow to further align the partial beams. The whole setup is in an ultrahigh vacuum at a pressure of \SI{5e-10}{\milli\bar} and shielded by a mumetal tube. 

\begin{figure*}
\includegraphics[width=7in]{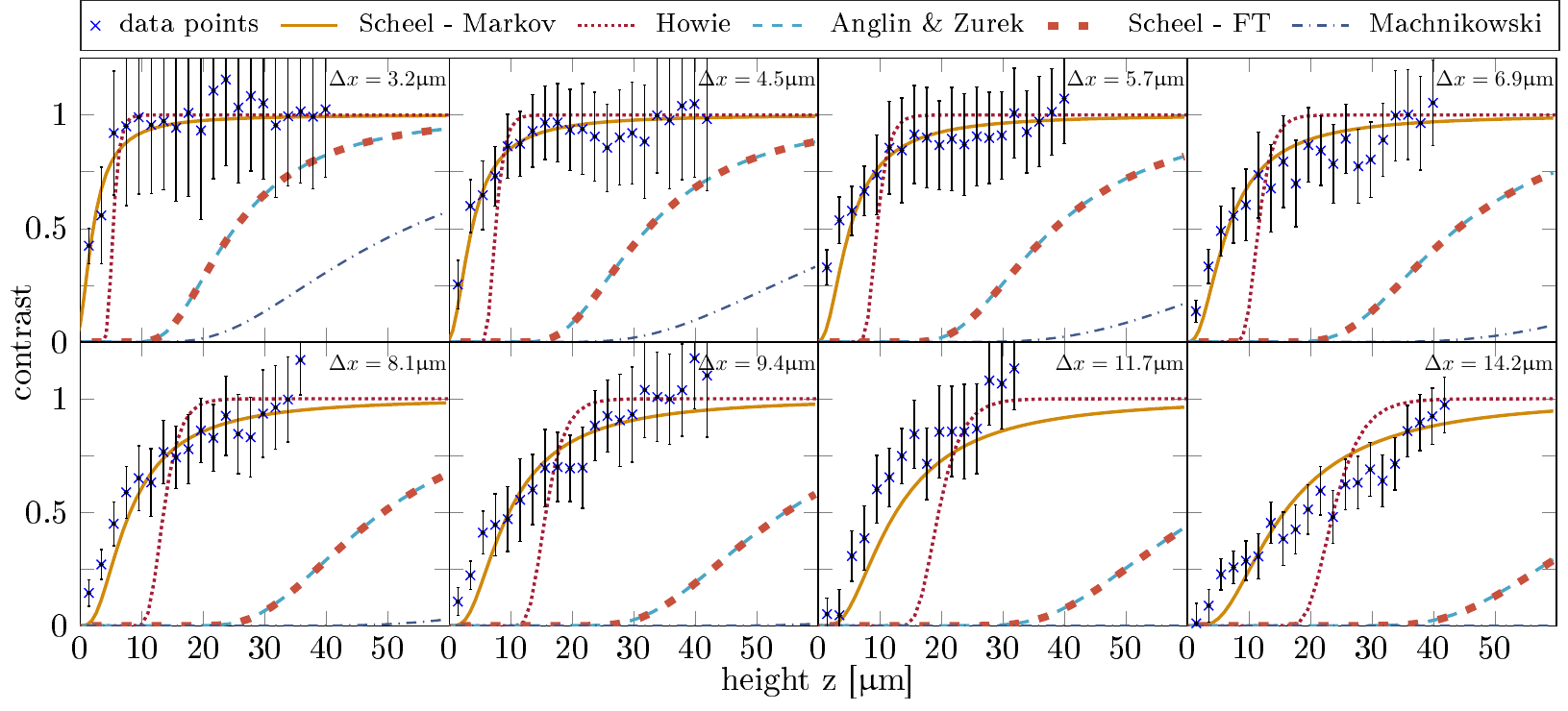} \caption{(Color online) Decoherence of electrons in a superposition state close to a doped silicon surface. The loss of visibility is plotted as a function of distance $z$ to the semiconducting plate for different beam path separations $\Delta x$ and compared to four current theoretical models \cite{Anglin1997,Scheel2012,Machnikowski2006,Howie2011}. } \label{fig2}
\end{figure*}

A correct identification of the beam path distance $\Delta x$ is crucial for the decoherence analysis. In the study by Sonnentag et al.~\cite{Sonnentag2007}, $\Delta x$ was calculated by transfer matrices. It cannot be determined from the fringe distance due to the unknown quadrupole magnification factor in this direction. In this experiment we present a novel method to directly measure $\Delta x$ with the Wien filter without destroying the wave function and independent of the magnification. 
As outlined in the literature \cite{Pooch2018,Nicklaus1993,Hasselbach2010}, the Wien filter introduces a longitudinal shift $\Delta y = \frac{L \, \Delta x \, U_{WF}}{2 \, D \, U_{beam}}$ between the spatially separated wave packets in the interferometer. Thereby, $L$ is the length of the Wien filter deflector plates, $\pm U_{WF}$ the applied voltage, $D$ is the distance between them and $e \, U_{beam}$ the electron energy \cite{Pooch2018}. The wave packets consist of the Fourier sum of linearly independent single electron plane waves. They have slightly different energies in accordance to the energy distribution of the beam source \cite{Nicklaus1993}. Changing $\Delta y$ reveals a Gaussian distributed dependency of the coherence contrast that defines the longitudinal coherence length of our beam: $l_c = \frac{2 \, U_{beam} \, \lambda}{\pi \, \Delta E}$, with $\lambda $ being the electron de Broglie wavelength \cite{Pooch2018,Nicklaus1993,Hasselbach2010}. We determined the energy width $\Delta E$ of our source in a separate measurement \cite{Pooch2018} where the beam paths were not additionally manipulated by the first quadrupole. The only separation was the biprism diameter which could be assessed in an electron microscope \cite{Schuetz2014}. A value $\Delta E =$ \SI[separate-uncertainty = true]{377(40)}{meV} was measured \cite{Pooch2018}, leading to $l_c =$ \SI[separate-uncertainty = true]{66(7)}{nm}. In the current setup, we could extrapolate $U_{WF}$ from the width of the Gaussian contrast curve determined when the wave packets are shifted by an amount of $\Delta y = l_c$ \cite{Pooch2018}. Inserting this in the equation above reveals a $\Delta x$ at the center of the Wien filter of \SI[separate-uncertainty = true]{2.9 \pm 0.4}{\micro\metre}. To check this result, we performed a calculation with transfer matrices for the optical elements in the beam path, leading to $\Delta x_{tm}=$ \SI{3.2}{\micro\metre} (as outlined in the supplemental material \cite{supplmat}). A beam path simulation with Simion (Sci.~Instr.~Serv., USA) yields $\Delta x_{Simion}=$ \SI{2.6}{\micro\metre}. Due to the good agreement, we consider the transfer matrices calculation a valid method and applied it in the data evaluation. 

\begin{figure*}
\includegraphics[width=7in]{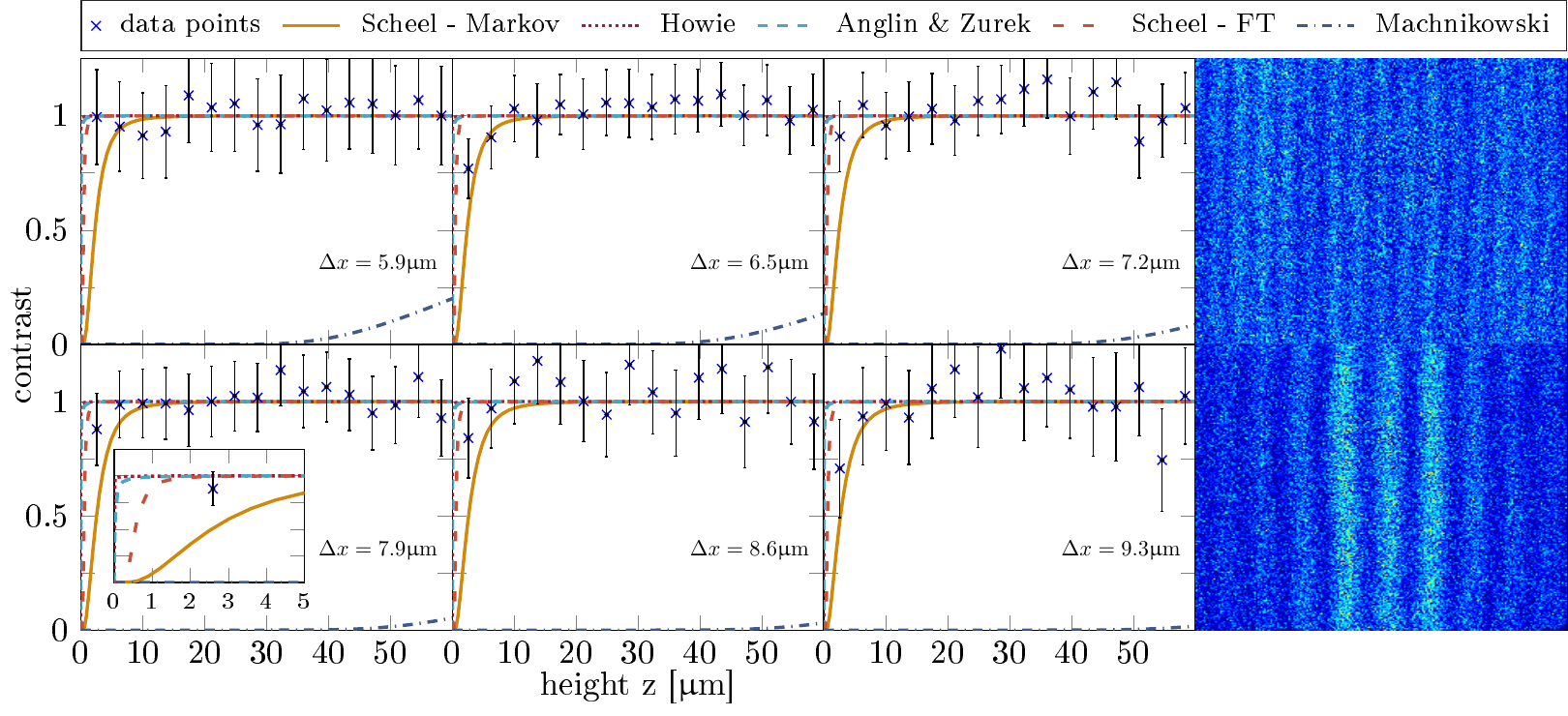} \caption{(Color online) Decoherence of the electron superposition state close to a gold surface. The effect is significantly smaller than with doped silicon due to the $\sim$10$^6$ lower resistivity. Two interference pattern for $\Delta x =$ \SI{9.3}{\micro\meter} and $\Delta x =$ \SI{6.5}{\micro\meter} are shown on the upper right and lower right side, respectively (bottom edge: gold surface, top edge: $z =$ \SI{80}{\micro\metre}).} \label{fig3}
\end{figure*}

{\it Results.--} For the decoherence analysis with the doped silicon surface, eight different beam path separations are chosen between $\Delta x =$ \SI{3.2}{\micro\metre} and \SI{14.2}{\micro\metre}. The resulting electron fringe distributions above the decoherence surface are illustrated in the Fig.~\ref{fig1} c). Each image contains \SI{500e3}{} electron counts. For contrast analysis, the interference patterns were insignificantly straightened by polynomial fits to decrease distortions. With decreasing beam-surface distance $z$, decoherence increases because of a stronger Coulomb interaction between the quantum system (electrons in superposition) and the decohering environment (semiconducting plate). A clear contrast loss is therefore observed for electrons passing close to the surface. As expected, the effect is stronger for larger beam path separations. With increasing $\Delta x$ the environment obtains more which-path information of the electrons, leading to stronger contrast loss and decoherence. For a better analysis, the interference contrast was determined as a function of the distance $z$ to the surface. The resulting distributions for the eight images in Fig.~\ref{fig1}~c) are shown in Fig.~\ref{fig2}. Each contrast point is determined by slicing a horizontal rectangular section of the image normal to the fringe orientation with a height of $\Delta z =$ \SI{2}{\micro\meter}. Then the counts therein were summed up to form a histogram that was fitted with the model function \mbox{$I(x)=I_0\cdot\left(1+C\cdot\cos(\frac{2\pi x}{s}+\phi_0)\right)\cdot {\rm sinc}^2(\frac{2\pi x}{s_1}+\phi_1)$} as described elsewhere \cite{Pooch2017,Pooch2018}. Thereby, $C$ is the interference contrast, $s$ the fringe distance, $\phi_0$ and $\phi_1$ phases, $I_0$ the average intensity and $s_1$ the width of the interference pattern. For contrast normalization, the average visibility of the upper \SI{5}{\micro\meter} in the data set was determined in each image separately and set to the contrast value equal 1. The data was compared to four current decoherence models \cite{Anglin1997,Scheel2012,Machnikowski2006,Howie2011} also plotted in Fig.~\ref{fig2} with a good overlap to the theory of Scheel et al.~in the Markov approximation \cite{Scheel2012}. The exact position of the surface at $z=0$ can only be determined within a small uncertainty. For the theoretical analysis it is set \SI{3}{\micro\meter} below the bottom edge in the images in Fig.~\ref{fig1} c). We describe the applied equations and different approximations in Fig.~\ref{fig2} and Fig.~\ref{fig3} in detail in the supplemental material \cite{supplmat}. 

We repeated the decoherence measurements with a gold surface of the same size. Due to its significantly lower resistivity, a smaller coherence loss is observed compared to doped silicon. The measured contrast versus surface distance data for different beam path separations are plotted in Fig.~\ref{fig3} together with two interference pattern for a large and small $\Delta x$. Again, we compared our data to the predicted decoherence distributions from four current theoretical models \cite{Anglin1997,Scheel2012,Machnikowski2006,Howie2011}. The results can clearly exclude the theory of Machnikowski \cite{Machnikowski2006}. The transition slopes between minimal and maximal contrast predicted by Howie \cite{Howie2011} (where the cut-off frequency was adapted with the conductibility of gold), Anglin \cite{Anglin1997} and Scheel in the finite temperature approximation (Scheel - FT) \cite{Scheel2012} are confined to a small region below $\sim$\SI{2}{\micro\metre}. This can be observed in the magnified section in the inset of Fig.~\ref{fig3}. For some path separations, our data has a better overlap with the theory of Scheel \cite{Scheel2012} in the Markov approximation similar to the measurements with the doped silicon surface. But the resolution in the sub-\SI{5}{\micro\metre} regime needs further improvement for a final conclusion which theory fits best for gold.

{\it Discussion.--} Several competing theoretical approaches exist to describe decoherence mediated by the Coulomb force, depending on different assumptions how the quantum state is modified by the environment. We compared four of them \cite{Anglin1997,Machnikowski2006,Howie2011,Scheel2012} to our experimental data with low and high conductibility surface materials (doped silicon and gold). Our measured decoherence dependence on $z$ and $\Delta x$ for silicon is in accordance to the experimental results from Sonnentag et al.~\cite{Sonnentag2007}. In that paper, two theories were fitted to the data that either included a fit factor with a large deviation to the predicted value (model from Anglin et al.~\cite{Anglin1997,Sonnentag2007}) or applied an approach that was originally only worked out for metals (model from Machnikowski \cite{Machnikowski2006}). Our results in Fig.~\ref{fig2} demonstrate that both theoretical approaches do not describe the decoherence well for silicon. Also a recently proposed decoherence model by Howie based on aloof scattering with long wavelength plasmons \cite{Howie2011,Beierle2018} does not reflect our experimental results. However, the data in this study clearly reveals that a recent theoretical model by Scheel et al.~\cite{Scheel2012,Karstens2014}, applying macroscopic quantum electrodynamics based on linear-response theory, describes the Coulomb-induced decoherence well without a fit factor. It is plotted in Fig.~\ref{fig2} in the Markov approximation \cite{Scheel2012} after a transformation of coordinates \cite{Karstens2014}. The finite temperature approximation \cite{Scheel2012} ("Scheel - FT" in Fig.~\ref{fig2}), however, does not fit to our outcome. The distribution overlaps with Anglin's approach \cite{Anglin1997}. The finite temperature approximation was also applied to the results in a recent decoherence study by Beierle et al.~\cite{Beierle2018}. They studied the decoherence of electrons in a quantum superposition state by the broadening of the diffraction peak aloof a doped silicon and a gold surface in a grating interferometer \cite{Beierle2018,Chen2018}. For the case of doped silicon, they come to the opposite conclusion by ruling out Scheel's theory while claiming to be in agreement with the data of Sonnentag et al.~\cite{Sonnentag2007}. However, they also acknowledged that Scheel's theory fits their data in case of a gold surface. 

Our study can clarify this situation by demonstrating that only a non-appropriate approximation was applied and Scheel's theory is correct in the less simplified Markov approximation. We solve the decoherence functional numerically for our experimental situation in the Markov approximation that assumes only the long-time limit \cite{Scheel2012}. It is closer to the full solution of the decoherence functional and significantly more complex to solve than the finite temperature approximation \cite{Scheel2012} (see suppl.~mat.~\cite{supplmat}). The high-temperature expansion in the finite temperature approximation requires additionally to fulfill the condition: $k_B T / \hbar \gg v/z$, with $v$ being the velocity of the electrons \cite{Scheel2012}. It can be discussed if the condition is sufficiently well met in our case and the experiment of Beierle et al.~\cite{Beierle2018}. This could be a possible reason for the deviation in the predictions from the two approximations. Our results for decoherence close to a gold surface are in accordance to the measurements by Beierle et al.~\cite{Beierle2018}. As it can be seen in Fig.~\ref{fig3}, the comparison between the different theories is not as clear as with silicon due to a much lower decoherence strength but also in this case, Scheel's theory \cite{Scheel2012} provides a good description within the error margins.

{\it Conclusion.--} Future hybride quantum devices with charged particles and novel quantum electron microscopic techniques rely on long coherence times of electrons in a superposition state close to a macroscopic conducting environment. The theoretical description of this Coulomb-induced decoherence was still an open question. This study tested the decoherence of aloof electrons in a superposition state close to a semiconducting and metallic surface by contrast loss in an electron biprism interferometer. Our results could exclude three current decoherence models \cite{Anglin1997,Machnikowski2006,Howie2011} and clearly favour a recent theory by Scheel et al.~\cite{Scheel2012} where decoherence is based on linear-response theory in macroscopic quantum electrodynamics.

{\it Acknowledgements.--} This work was supported by the Deutsche Forschungsgemeinschaft (DFG) through the research grant STI 615/3-1. Work at the Molecular Foundry was supported by the Office of Science, Office of Basic Energy Sciences, of the U.S. Department of Energy under Contract No.~DE-AC02-05CH11231. We also acknowledge support from the Vector Stiftung. We thank Colin Ophus for the fruitful discussions.

\newpage

\section*{Supplemental Material}

\subsection{Applied decoherence models} 

In the analysis of our decoherence measurements of an electronic superposition state close to a semiconducting and metallic surface, we compared four theoretical models. Here, we point out briefly the equations from the according publications  \cite{Scheel2012,Karstens2014,Anglin1997,Sonnentag2007,Machnikowski2006,Howie2011} that were applied for the theoretical curves in fig.~2 and fig.~3 and refer to the original literature for a detailed description of the parameters. 

The best agreement to our experimental data was achieved with the model from Scheel et al.~\cite{Scheel2012}. In their article they formulated the general equation for the decoherence functional in equ.~(11). Since it is not analytically tractable, several levels of approximation were made for simplification. The minimum assumption was the long-time, slow velocity limit. Thereby, the interaction time with the electrons (time the electrons need to pass the plate) is long in regard to other timescales in the system. The resulting expression for the decoherence functional is the {\bf Markov approximation} (equ.~(15) in \cite{Scheel2012}):
\begin{align}
    \Gamma[c] &= -t \frac{q^2}{\epsilon_0 \hbar}\int\int\frac{k_\parallel \text{d}k_\parallel\text{d}\varphi}{(2\pi)^2}[1-\cos(k_\parallel \Delta x \sin\varphi)]\\ \nonumber
    \times&[2\bar{n}_{th}(\eta+1)]\frac{e^{-2k_\parallel z \gamma(\varphi)}}{2k_\parallel\gamma(\varphi)}\\ \nonumber \times&\text{Im}\left\{r_p(k_\parallel,\eta)-
    \frac{v^2}{c^2}[r_p(k_\parallel,\eta)\cos^2\varphi-r_s(k_\parallel,\eta)\sin^2\varphi]\right\}
\end{align}
with $\eta = k_\parallel v |\cos\varphi|$ and $\Delta x$ the beam path separation.

The effective contrast is calculated as $V=e^{-\Gamma[c]}$. In fig.~2 and fig.~3, this equation is solved numerically in its equivalent form after a transformation in Cartesian coordinates as outlined in \cite{Karstens2014}. The integral was numerically solved by replacing it with sums. Because of the different components of the function, an adaptive mesh was used.\\

As a next step for easier calculation of the decoherence functional, a high-temperature expansion is performed in connection to the Drude model. In Scheel et al.~it is called the {\bf finite temperature approximation} (equ.~(17) in \cite{Scheel2012}):

\begin{equation}
\Gamma_{th}[c] = -t\frac{q^2 k_B Tr_p^{'(0)}}{2 \pi \epsilon_0\hbar^2}\left[\frac{1}{2z}-\frac{1}{\sqrt{(2z)^2+L^2}}\right]
\end{equation}

where $L$ is the length of the decoherence surface.\\

Another theory that was applied on our data is by Anglin and Zurek \cite{Anglin1997} as expressed in \cite{Sonnentag2007}:
\begin{equation}
    \Gamma[c]=\frac{\pi e^2 k_bT \rho L (\Delta x)^2}{4 h^2 v z^3}
\end{equation}
with $\rho$ the resistance of the material. \\

The third theory is by Machnikowski (equ.~(22) in \cite{Machnikowski2006}):
\begin{equation}
    \frac{1}{\lambda} = \frac{k_B T}{2\pi^2\hbar v}\gamma\left( \frac{\Delta x}{z}\right)\mu\left(\frac{m_{eff}e^2}{2\pi\epsilon_0\epsilon_i\hbar^2 k_f}\right)
\end{equation}
with $m_{eff}$ the effective electron mass, $k_F$ the Fermi wave vector. It includes the "geometric" function
\begin{equation}
    \gamma(\xi) = \frac{1}{2}\int\limits_{-\infty}^\infty \frac{1}{1+u^2}\ln\left(1+\frac{\xi^2}{4}\frac{u^2}{1+u^2}\right)\text{d}u
\end{equation}
and the "material" function 
\begin{equation}
    \mu(\zeta)=\frac{\zeta^2}{4}\int\limits_0^1\frac{1}{u^3}\left(1+\frac{\zeta}{4\pi u^2}\left(1+\frac{1-u^2}{2u}\ln\left(\frac{1+u}{1-u}\right)\right)\right)^{-2}\text{d}u
\end{equation}
The effective contrast is calculated as $V = e^{-\frac{L}{\lambda}}$.\\
\\
The forth theory is by Howie \cite{Howie2011}:
\begin{align}
    P\left(\frac{z}{\alpha\Delta x}\right)=&\left[\frac{e^2 L {\omega_m}^2}{4\pi^2\hbar\sigma v^2}\right]\int\limits_{\alpha/\delta x}^\infty \frac{exp(-2zq_z)}{q_z}\text{d}q_z\\
    =& \left[\frac{e^2 L {\omega_m}^2}{4\pi^2\hbar\sigma v^2}\right] E_1\left(\frac{2\alpha z}{\Delta x}\right)
\end{align}
where $\omega_m$ is the cut-off frequency of the dielectric response function, $\sigma$ the conductivity of the surface and 
\begin{equation}
    E_1(\eta)=\int\limits_\eta^\infty \frac{exp(-x)}{x}\text{d}x.
\end{equation}
In \cite{Howie2011}, the parameter $\alpha$ is set as $\alpha = 2$, the fringe visibility is given as $V=e^{-P}$.\\
In the data evaluation of the decoherence above doped silicon in Fig.~2, all theory curves were shifted by \SI{-3}{\micro\meter} due to a small uncertainty in the determination of the exact position of the surface at \mbox{$z=0$}.

\subsection{Applied transfer matrices to calculate the beam path separation}

The beam path separation $\Delta x$ is a crucial parameter in this experiment for the correct determination of the decoherence. We measured the beam path directly with a novel method using the Wien filter as described in the main paper. It is a time consuming process that could not be performed between each decoherence measurement with a different beam path separation due to long term stability of the whole setup. For that reason, the Wien filter method was applied once to verify the correctness of the beam path separation calculated by transfer matrices. The matrices for each major component are defined in the following expressions \cite{Wollnik1987,Maier1997} and applied successively according to the setup in fig.~1~a).

The beam path starts with an incoming vector $v_1$ with an initial position $x_0$ with respect to the optical axis and an initial angle $\alpha_0$:
\begin{equation}
v_1 = 
\begin{pmatrix}
x_0\\
\tan\alpha_0
\end{pmatrix}
\end{equation}

A drift distance $d$ between different components is described by the matrix:

\begin{equation}
M_{drift} = 
\begin{pmatrix}
1 & d \\
0 & 1
\end{pmatrix}
\end{equation}

The biprism is in principle a kind of deflector with different deflection angles depending on which side the electron passes: 
\begin{equation}
M_{BP} = 
\begin{pmatrix}
1 & 0 \\
0 & \frac{\tan(\alpha_0\pm\gamma)}{\tan\alpha_0}
\end{pmatrix}
\end{equation}
with
\begin{equation}
\gamma = \frac{\pi}{2\cdot\log(\frac{r_g}{r_{BP}})}\cdot\frac{U_{BP}}{U_{beam}} 
\end{equation}
where $r_g$ is the distance between the biprism wire and the grounded plates, $r_{BP}$ is the radius of the biprism wire, $U_{BP}$ the applied voltage on the wire and $e\,U_{beam}$ the electron energy.

Every quadrupole has a focusing and a defocusing plane. The relevant beam path separation in our setup is along the optical axis such as illustrated in fig.~1 b). For beam focusing and fringe pattern magnification in different sections, the first and the forth quadrupole lenses are in reverse polarity than the second and third quadrupole lenses. The corresponding transfer matrices are:

\begin{align}
M_{QP focusing} &= 
\begin{pmatrix}
\cosh(kl) & \frac{1}{k}\sinh(kl)\\
k\sinh(kl) & \cosh(kl)
\end{pmatrix}\\
M_{QP defocusing} &=
\begin{pmatrix}
\cos(kl) & \frac{1}{k}\sin(kl)\\
-k\sin(kl) & \cos(kl)
\end{pmatrix}
\end{align}
\\
with $k=\sqrt{\frac{U_q}{{g_0}^2U_{beam}}}$, the applied voltage $U_q$ on the quadrupole electrodes and the inner diameter $g_0$ of the quadrupole. All remaining electron optical elements in the setup are not required to calculate the beam path separation.

\end{document}